\documentclass{article}

\usepackage{arxiv}
\usepackage[utf8]{inputenc} 
\usepackage[T1]{fontenc}    
\usepackage{hyperref}       
\usepackage{url}            
\usepackage{booktabs}       
\usepackage{amsfonts}       
\usepackage{nicefrac}       
\usepackage{microtype}      
\usepackage{graphicx}
\graphicspath{ {./Figures/} }

\usepackage{array}
\usepackage{tabularx}
\usepackage{amsmath}
\usepackage{amssymb}
\usepackage{enumitem}
\usepackage{float}
\usepackage[numbers]{natbib} 
\usepackage{caption}
\usepackage[table]{xcolor}
\newcommand{\sym}[1]{\ensuremath{^{#1}}}
\usepackage{tcolorbox}
\tcbuselibrary{skins}
\tcbuselibrary{breakable}
\usepackage{listings}
\usepackage{xcolor}

\tcbset{
  mybox/.style={
    enhanced,
    colback=gray!5,
    colframe=gray!50!black,
    fonttitle=\bfseries,
    boxrule=0.5pt,
    arc=2pt,
    left=5pt,
    right=5pt,
    top=5pt,
    bottom=5pt,
  }
}

\date{}

\makeatletter
\renewcommand{\@maketitle}{%
  \newpage
  \null
  \vskip 2em%
  \begin{center}%
  \let \footnote \thanks
  {\LARGE \@title \par}%
  \vskip 1.5em%
  {\large
    \lineskip .5em%
    \begin{tabular}[t]{c}%
      \@author
    \end{tabular}\par}%
  \vskip 1em%
  \end{center}%
  \par
  \vskip 1.5em}
\makeatother

\title{Architectural Vulnerability and Reliability Challenges in AI Text Annotation: A Survey-Inspired Framework with Independent Probability Assessment}
\author{Linzhuo Li}
\begin{document}

\maketitle

\begin{abstract}
Large Language Models, despite their power, have a fundamental architectural vulnerability stemming from their causal transformer design -- order sensitivity. This architectural constraint may distorts classification outcomes when prompt elements like label options are reordered, revealing a theoretical gap between accuracy metrics and true model reliability. The paper conceptualizes this vulnerability through the lens of survey methodology, where respondent biases parallel LLM positional dependencies. Empirical evidence using the F1000 biomedical dataset across three scales of LLaMA-3.1 models (8B, 70B, 405B) demonstrates that these architectural constraints produce inconsistent annotations under controlled perturbations. The paper advances a practical solution for social science -- Independent Probability Assessment --which decouples label evaluation to circumvent positional bias inherent in sequential processing. This approach yields an information-theoretic reliability measure (R-score) that quantifies annotation robustness at the case level. The findings establish that architectural vulnerabilities in causal transformers require methodological innovations beyond accuracy metrics to ensure valid social science inference, as demonstrated through downstream regression analyses where order-sensitive annotations significantly alter substantive conclusions about scientific impact.

\end{abstract}


\section{Introduction}

Large Language Models (LLMs) have become indispensable for large-scale text classification and annotation in social sciences, enabling novel investigations of massive corpora \citep{tan2024large, yan2023cipta, bail2024can}. Yet, despite their capacity to handle subtle nuances, researchers face a persistent challenge: \textbf{reliability}(\citep{palmer2024using,rytting2023towards, zhang2023safetybench,tornberg2024best, ziems2024can, dentella2023systematic,adewumi2024limitations,bisbee2024synthetic}). Traditional evaluation often relies on comparing model predictions against human expert annotations, yielding an accuracy score. However, focusing exclusively on accuracy can obscure fragile or inconsistent reasoning processes, especially for rare or conceptually intricate categories \citep{tornberg2024best, adewumi2024limitations}.

A growing body of work shows that LLM outputs not just contain biases\citep{caliskan2017semantics}, but can also be sensitive to seemingly minor changes in prompts, including reordering the list of possible labels or adjusting the exact phrasing \citep{rytting2023towards, pezeshkpour2023large}. This phenomenon is especially pertinent for causal, decoder-only transformer architectures. Because they process text sequentially, with each token attending only to prior tokens, small variations in token or label order can lead to significant shifts in predicted outcomes. The resulting \emph{order sensitivity} introduces a reliability gap not captured by simple accuracy metrics.

In other words, a model can produce ``correct'' annotations for a given prompt design, yet be highly inconsistent when the same question is posed with minor structural changes. For social scientists, whose interest often lies in stable, theoretically grounded coding of phenomena, this poses a fundamental challenge: \emph{how can one trust that an LLM annotation is robust once it leaves the controlled conditions of a benchmark prompt?} Furthermore, how can researchers detect, at the case level, which annotations are most vulnerable to these artifacts to ensure construct validity \citep{strauss2009construct}? 

To address these issues, this paper proposes and demonstrates a two-part approach:

\begin{enumerate}
\item \textbf{Diagnosing order sensitivity with survey-inspired interventions.} Drawing on principles akin to survey design, the research introduces controlled perturbations to test whether annotations remain stable under changes in label ordering, question structure, and reverse-coded logic. These interventions produce high ``flip rates'' in many cases, thereby revealing the hidden brittleness inherent to causal transformers' sequential attention.
\item \textbf{Mitigating causal ordering bias and quantifying reliability with independent probability assessment.} As a primary technical solution, the paper proposes an \emph{independent} querying strategy for classification tasks. Instead of providing a single prompt containing all possible labels (which the causal model processes in an order-sensitive fashion), each label is queried separately. Normalizing these independently computed probabilities yields a distribution free from positional bias. Building on this unbiased distribution, the study introduces an information-based \textit{R-score} to measure the degree to which the model's predicted distribution diverges from random guessing. This per-case metric identifies which annotations are genuinely confident and which remain unreliable.
\end{enumerate}

The framework is evaluated on annotating 816 scientific abstracts from the F1000 dataset, using three sizes of LLaMA-3.1 models (8B, 70B, 405B parameters). Despite larger models achieving higher accuracy, non-trivial proportions of annotations change (or ``flip'') under small prompt order variations, highlighting that accuracy alone overlooks important reliability pitfalls. Moreover, by comparing the intervention-based diagnostics to the new R-score, the study shows that cases with lower R-scores indeed exhibit higher flip rates, affirming that the R-score captures genuine reliability. Finally, a downstream regression example demonstrates that ignoring these reliability factors can lead to unstable or even contradictory findings about scientific impact.

Overall, the contributions of this paper are:

\begin{itemize}[leftmargin=*]
\item \textbf{Identify the causal transformer's order-sensitivity problem in annotation tasks.} The work clarifies how conventional multi-choice prompts exacerbate this problem and why standard accuracy metrics fail to detect it.
\item \textbf{Introduce survey-inspired perturbation interventions} (option randomization, prompt reordering, reverse validation) as an experimental toolkit to diagnose and reveal reliability risks due to the sequential nature of LLMs.
\item \textbf{Propose a novel independent probability assessment strategy} that disentangles label queries, mitigating positional bias and yielding a more faithful view of the model's internal distribution.
\item \textbf{Develop an information-based R-score} to quantify case-level annotation reliability, with empirical tests linking low R-score to high intervention-induced flips.
\item \textbf{Demonstrate consequential impacts in downstream analyses}, warning social scientists and applied researchers that ignoring these reliability methods can yield misleading empirical conclusions.
\end{itemize}

The remainder of the paper is structured as follows:
Section \ref{sec:causal-transformer} details the causal attention mechanism and explains how it introduces order sensitivity.
Section \ref{sec:framework} presents the reliability assessment framework and discusses both the survey-inspired diagnostics and the independent probability assessment (including the R-score).
Section \ref{sec:data-methods} describes the dataset, models, and methodological implementation.
Section \ref{sec:results} reports empirical findings on flip rates, reliability distributions, and downstream regressions.
Section \ref{sec:discussion} offers a broader discussion of implications, limitations, and future directions.
Section \ref{sec:conclusion} concludes.

\section{Architectural Vulnerability of LLM}
\label{sec:causal-transformer}

Building on the foundational insight from quantitative text analysis that "all quantitative models of language are wrong--but some are useful" \citep{grimmer2013text}, researchers have long recognized the inherent simplifications and assumptions made by computational approaches when analyzing complex human language. Early methods, from simple word frequency counts \citep{franzosi2004words} to topic models \citep{blei2003latent}, were understood as useful approximations of meaning, operating on simplified representations of text. The advent of Large Language Models (LLMs), particularly the causal transformer architectures that dominate current applications, appears to represent a significant leap, offering seemingly more sophisticated and fine-grained understandings of language itself. However, while these models capture linguistic complexities far beyond their predecessors, they do not escape the fundamental "wrongness" inherent to all computational approaches to language -- they simply manifest it in novel ways.

Large Language Models' causal, decoder-only transformer architectures impose specific structural constraints that significantly shape how they engage with and categorize textual data. Most critically, when applied to tasks structured as ordered lists (like multi-choice classification prompts), these models exhibit a problematic order sensitivity, leading to predictions contingent on arbitrary positional cues rather than a stable understanding of content relative to categories. This architectural vulnerability is not merely a technical limitation but functions akin to methodological choices or social structures, framing possibilities while simultaneously imposing constraints and potential biases. For social scientists leveraging these models for annotation, recognizing and accounting for these structural predispositions is crucial to ensure that LLM-driven annotation pipelines produce not just seemingly accurate, but truly reliable data for social scientific inquiry.

\subsection{Causal Attention as a Structural Constraint}
The core mechanism of causal transformers is \emph{sequential processing} via causal attention \citep{floridi2020gpt}. Each token in the input sequence is processed in order, and its representation is built by attending only to the tokens that preceded it. In other words, tokens before cannot see tokens after, and information only flows one way. This architectural design appears innocuous on the surface--even advantageous for its primary purpose of generating coherent text token by token. However, when these same models are applied to classification tasks presented sequentially (often as a list of choices) in prompts, this seemingly neutral architecture introduces a fundamental structural constraint: the model's interpretation and scoring of a given classification option become dependent on its position within the input sequence relative to other options. What initially presents as a benign technical implementation reveals itself as a consequential limitation that can significantly distort classification outcomes in ways invisible to casual users.

Consider a typical multi-choice prompt format:
\begin{tcolorbox}[mybox, title=Multi-Choice Prompt Example]
\textbf{Prompt:} Given the abstract: [\emph{abstract text}].
What is the main contribution type of this paper?
(A) Option 1
(B) Option 2
(C) Option 3
\textbf{Model Output:} (B)
\end{tcolorbox}

\begin{figure}[H]
\centering
\includegraphics[width=1\textwidth]{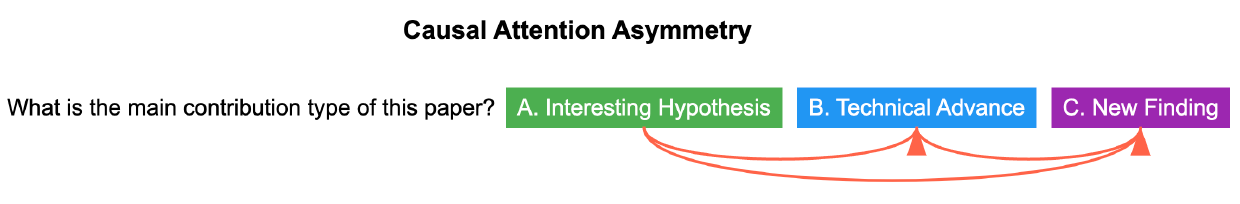}
\caption{The Causal Asymmetry of Information Accumulation in LLM}
\label{fig:figcausal}
\end{figure}

Under a causal attention mechanism, the model reads "Option 1" (A), then processes "Option 2" (B) while attending to "Option 1," and finally processes "Option 3" (C) while attending to both "Option 1" and "Option 2." (see Figure \ref{fig:figcausal}) The internal state and probability assigned to "Option 3" are thus causally influenced by the prior appearance and processing of "Option 1" and "Option 2." Altering the sequence of options -- swapping (A) and (B), for instance -- changes the causal history for downstream tokens, leading to different internal representations and potentially altering the model's final predicted label \citep{pezeshkpour2023large}. This \emph{order sensitivity} is not a random error; it is a systematic artifact stemming from the causal architecture's sequential processing when confronted with a task structure (multi-choice list) that ideally requires order-invariant evaluation. Thus, the model isn't actually evaluating the options independently.

\subsection{Non-Causal Architectures}
To highlight this specific vulnerability, it's useful to briefly contrast with non-causal transformer architectures, such as bidirectional encoders (e.g., BERT, RoBERTa) \citep{devlin2019bert,liu2019roberta} or encoder-decoder models often used for classification. These models typically process the input text to be classified using attention mechanisms that consider the entire sequence bidirectionally. In a standard fine-tuning setup for classification, the list of possible categories usually defines the output layer's structure, not elements processed sequentially within the encoder's main input sequence.
In such standard non-causal classification pipelines, the model forms a representation of the input text based on its holistic structure, and then a classification head maps this representation onto the predefined categories. While these models have their own sensitivities and biases (e.g., stemming from pre-training data or fine-tuning specifics), they are generally \emph{less susceptible to the specific type of order sensitivity} induced by the sequential processing of response options in the prompt containing multiple choices.

\subsection{Algorithmic Structure and Methodological Implications}
The prominence of causal LLMs in annotation, despite this order sensitivity, stems from their remarkable flexibility in zero-shot and few-shot learning. Social scientists are drawn to their ability to perform new tasks based on simple prompts, bypassing the need for large labeled datasets and task-specific model fine-tuning \citep{tan2024large, yan2023cipta,von2024vox, argyle2023out}.

The paradox emerges when considering why causal transformer architectures have nonetheless become dominant in contemporary applications. Their generative capacities--the ability to produce coherent, contextually relevant text that appears almost human-like--are precisely what makes them so powerful and appealing across domains. This generative prowess stems directly from the causal attention mechanism that creates meaningful sequential dependencies between tokens. The very architectural feature enabling GPT models to generate compelling narratives, explanations, and arguments is simultaneously the source of their vulnerability when applied to classification tasks requiring position-invariant evaluation of options.

This tension reflects a fundamental principle in social science methodology, where distinct research traditions or 'cultures' are shaped by differing goals and practices, leading to inherent trade-offs in their respective strengths and limitations \citep{mahoney2006tale}. Viewed through this lens, these trade-offs are evident across methodologies: for instance, qualitative traditions, often oriented toward explaining specific outcomes, offer rich contextual understanding but may sacrifice generalizability; experimental designs, focused on estimating average treatment effects, provide causal precision but often lack ecological validity. Similarly, causal transformer architectures offer remarkable generative capabilities while introducing structural biases in certain analytical contexts. The difference, critically, is that while methodological limitations in traditional social science are generally explicitly acknowledged and accounted for, the architectural constraints of LLMs often operate invisibly and become tacit knowledge of this technology \citep{mackenzie1995tacit}, masked by the apparent sophistication and fluency of model outputs.

From a critical algorithm studies perspective, this architectural constraint becomes a methodological structure that can inadvertently impose a particular, potentially distorting, logic onto the data. The model isn't neutrally identifying inherent categories; its process of categorization is influenced by the arbitrary sequence in which options are presented. This resonates with critiques arguing that algorithms don't just reflect social reality but can actively construct it by imposing predefined categories or processing logics that privilege certain outcomes or perspectives \citep{noble2018algorithms} . In this case, the causal structure can lead the model to "see" patterns or distinctions (or fail to see them) based on positional cues rather than solely on the substantive content relative to the definition of each category.

\subsection{Limitations of Accuracy-Centric Evaluation}

When evaluating AI text annotation performance, researchers commonly rely on a simple accuracy metric, comparing the model's predicted label to a "ground truth" annotation. However, this evaluation approach suffers from significant limitations that fail to capture critical vulnerabilities in model responses, particularly for complex social science text classification tasks. Accuracy metrics overlook prompt-dependence --where a model might provide an "accurate" answer for a specific label ordering, yet produce entirely different results when those same options are merely rearranged. Additionally, single top-label predictions conceal the model's uncertainty levels, making it impossible to distinguish between predictions based on strong evidence versus those essentially "guessing" among seemingly plausible answers. Moreover, high overall accuracy can mask instabilities in the model's handling of rare categories or conceptually challenging classifications \citep{tornberg2024best}.

\begin{figure}[H]
\centering
\includegraphics[width=0.35\textwidth]{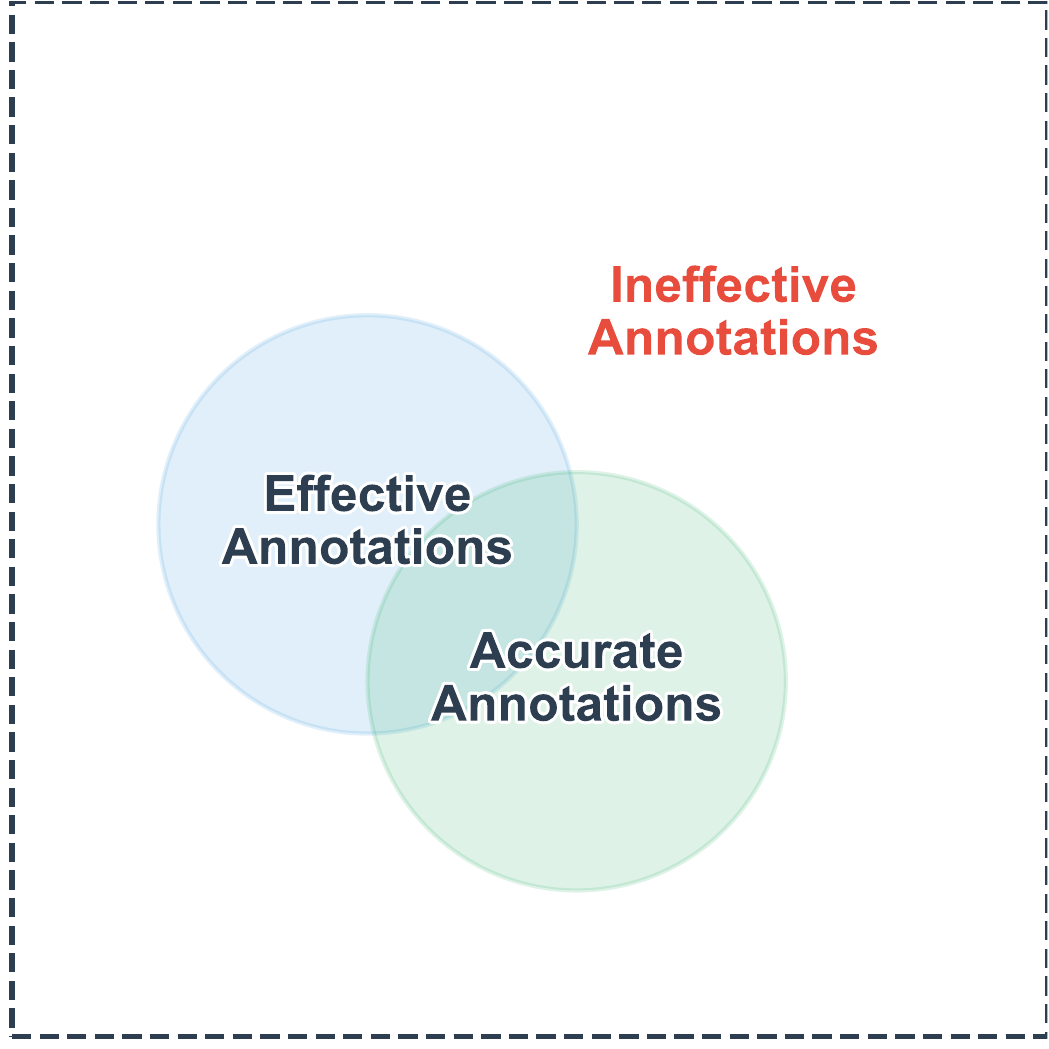}
\caption{Effective and ineffective AI annotations}
\label{fig:fig1}
\end{figure}

As illustrated in Figure \ref{fig:fig1}, there exists a conceptual distinction between the boundary of effectiveness in Large Language Models text annotation and the traditionally emphasized accuracy metrics. While previous studies have typically adopted accuracy-based evaluations, this paper argues that LLM reliability cannot be assessed through accuracy metrics alone. Accuracy, typically measured by comparing LLM outputs with expert annotations, captures only one dimension of reliability. This situation parallels challenges in survey methodology, where respondents may provide seemingly "correct" responses that nonetheless prove ineffective. Similarly, LLM annotations require evaluation for their substantive effectiveness. The figure depicts two partially overlapping circles representing accurate and effective annotations, with their partial overlap illustrating this distinction--accurate annotations may match expert coding but still fail to capture the underlying construct that researchers truly care about.

If the goal is to rely on LLM annotations for downstream analyses (such as measuring concept prevalence in large textual corpora or investigating relationships between that concept and outcomes of interest), being "correct" in an absolute sense proves insufficient. The annotation must also demonstrate robustness to minor changes in conditions. Without assessing how brittle these annotations may be, social scientists risk building entire analytical frameworks on unstable foundations, potentially compromising the reliability of their research conclusions.

Therefore, a critical methodological challenge in using causal LLMs for annotation is to move beyond aggregate accuracy metrics and develop strategies to: 1) diagnose the extent to which this architectural vulnerability translates into order-sensitive annotations for a given task and model; and 2) devise methods to elicit the model's preferences in a way that is robust to these structural biases, thereby obtaining a more reliable measure of its confidence in assigning a category. The framework presented in the following section addresses these challenges directly.

\section{Framework for Assessing and Quantifying Reliability}
\label{sec:framework}

In this section, this paper presents a two-part framework designed to (1) diagnose order sensitivity in causal LLM annotations and (2) mitigate it via an alternate query strategy that allows for calculating a case-level reliability measure.

\subsection{Part I: Diagnosing Order Sensitivity via Survey-Inspired Interventions}
While recent literature explores using LLMs to simulate human survey respondents (\citep{von2024vox, argyle2023out, bisbee2024synthetic, kozlowski2024silico}), this paper reverses the direction of knowledge transfer and argues that survey methodology's rich tradition of diagnosing and improving response reliability offers a framework for addressing LLM annotation challenges.

Despite the apparent differences between survey respondents and Large Language Models (LLMs), both can exhibit behavior that does not necessarily stem from deep engagement or true "understanding." Survey research has documented how participants sometimes take cognitive shortcuts--known as satisficing \citep{krosnick1991response}, \citep{barge2012using}. Proposed by \citep{simon1957behavioral} in a general sense and by \citep{krosnick1991response} specifically in survery research, satisficing describes how respondents may skip or abbreviate the four cognitive processes of comprehension, retrieval, judgment, and response selection to conserve effort. Instead of formulating a fully reasoned answer, they opt for a response that seems good enough. Sometimes, satisfying can also take a more extreme form as in the case of careless responding \citep{meade2012identifying,johnson2005ascertaining, ward2023dealing}: not only are participants skipping deeper thought, but they may be ignoring the survey content altogether.

The framework adapts three perturbation'' strategies from survey methodology to reveal whether and how much a model is sensitive to small prompt changes. These strategies---\emph{option randomization}, \emph{position randomization}, and \emph{reverse validation}---can be understood as  "screener question" \citep{berinsky2014separating}, and mimic how survey researchers detect inattentive or satisficing'' respondents by shuffling question order or wording \citep{krosnick1991response, meade2012identifying}.

\begin{enumerate}[label=\Alph*)]
\item \textbf{Option Randomization.} Vary the order of category labels (e.g., (A) New Finding, (B) Technical Advance, (C) Interesting Hypothesis) across different prompts for the same text. If a model frequently changes its predicted label when the category order changes, that signals reliance on positional cues rather than content.
\item \textbf{Position Randomization.} Alter the structure or location of the question in the prompt. For instance, moving the classification question from the end to the beginning or around different segments. Significant changes in output reveal instability tied to these structural positions.
\item \textbf{Reverse Validation.} Present an inverted or negated version of the classification. For example, after asking Which category best describes this paper?'' also ask Which category does \emph{not} describe this paper?'' If the model's answers are inconsistent, this indicates that it does not genuinely grasp the underlying text meaning and is instead responding to superficial cues.
\end{enumerate}

\paragraph{Flip Rate as Diagnostic Metric.}
For each of these interventions, the study measures the proportion of cases where the model's top choice \emph{flips} compared to its original assignment. A high flip rate indicates that the model is highly sensitive to order or structural changes, suggesting low reliability. Conversely, a low flip rate suggests the model is more robust.

\subsection{Part II: Mitigating Order Bias via Independent Probability Assessment}
While the above interventions reveal the extent of a model's order sensitivity, the paper also offers a \emph{technical approach} to mitigate the causal transformer's positional bias in multi-choice classification. This is accomplished by disentangling each label's evaluation from the context of other labels.

\subsubsection{Independent Binary Queries}
Instead of presenting all possible categories in one prompt, the approach breaks the task into multiple \emph{binary} queries as follows:

\begin{tcolorbox}[mybox, title=Independent Probability Assessment -- Binary Query]
\textbf{Example Prompt for Category C: Are we dealing with a Technical Advance''?}\\ Given the abstract: [\emph{abstract text}].   Please answer Yes'' or ``No'' only.
\end{tcolorbox}

This is done separately for each category that needs to be assessed. Because the model now sees only one category at a time, it cannot exploit relative ordering or rely on sequence-level heuristics that compare multiple labeled options. Each query yields a probability of ``Yes'' (i.e., the category applies). By repeating for all categories, this yields independent probabilities:
\[
p(\text{Yes} \mid \text{category } i), \quad \forall i \in \{\text{A, B, C}\}.
\]

In a single-label classification context (only one category can apply), we can normalize these probabilities across all categories to obtain:
\[
p(c_i) \;=\; \frac{p(\text{Yes} \mid c_i)}{\sum_j p(\text{Yes} \mid c_j)}.
\]
This approach bypasses the causal attention trap in multi-choice prompts, because the model does not read or process the other categories at the same time.

By repeating this process for each category ("Interesting Hypothesis", "Technical Advance", and "New Finding"), we can obtain a comprehensive distribution that represents the model's genuine preferences without positional biases. It can handle both exclusive classifications (where exactly one category applies) and non-exclusive scenarios (where multiple categories may apply simultaneously). This approach can be understood as the LLM-version of projecting meanings onto specific dimensions (\citep{kozlowski2019geometry,rodriguez2022word,charlesworth2022historical}) —thus representing a "geometry of thinking" where each query projects the input text onto a distinct semantic axis, albeit using natural language. This controls for information asymmetry, allowing us to capture the model's genuine preference distribution.

\subsubsection{Information-Based Reliability Score (R-score)}
Although this independent probability assessment helps us avoid ordering bias, we still need a method to distinguish whether the model holds a \emph{strong} preference for one category or is nearly guessing. The study thus defines a per-case \emph{R-score} measuring how far the model’s final distribution is from uniform:

\begin{tcolorbox}[mybox, title=R-score Definition]
\[
R = D_{KL}\bigl(P \,\|\, U\bigr) \;=\; \sum_{i=1}^{k} p_i \log \!\Bigl( \frac{p_i}{1/k} \Bigr),
\]
\end{tcolorbox}

where $P = \{p_1, p_2, \dots, p_k\}$ is the normalized distribution over $k$ categories (each derived via independent queries), and $U=\{1/k,\dots,1/k\}$ is the uniform distribution. If $R$ is near zero, the model is effectively distributing probability evenly across all labels, indicating high \emph{uncertainty} (i.e., the model is close to random guessing). A high $R$ indicates the model strongly favors one category over the others.

In practice, similar to the choice of p-values in statistical inference,the thresholds of $R$-score were empirically selected based on intuitive probability distributions for a three-option classification scenario. A distribution close to uniform (low R-score) suggests the model cannot meaningfully differentiate between categories, while a highly skewed distribution (high R-score) indicates strong preference for a particular category. The threshold of KL = 0.06 corresponds approximately to a distribution of [0.5, 0.25, 0.25], where the top probability is equal to the sum of remaining options. KL = 0.36 corresponds to a distribution of [0.75, 0.125, 0.125], where the top probability is three times of the rest others, demonstrating a clearer model preference. KL = 0.7 corresponds to a distribution of [0.9, 0.05, 0.05], representing a case where the model shows very strong confidence (90\% of the time) in its top prediction. In other words, if the model repeatedly annotates this case randomly, the result is consistent with the "top" label in 90\% cases. Critically, the R-score is calculated at the \emph{case level}, so we can identify exactly which instances are robust and which are not. 

\subsection{Linking the Two Parts:}
While the survey-inspired interventions \emph{detect} order sensitivity, the independent assessment \emph{corrects} for it when computing a probability distribution. Cases that exhibit high flip rates under the interventions generally turn out to have low R-scores in the independent assessment, confirming that the same underlying weakness is being measured. For social science applications, researchers can use these diagnostics jointly: 
(1) run interventions to measure how sensitive each annotation is, 
(2) re-annotate crucial tasks using independent queries for robust probability estimates, and 
(3) filter or prioritize expert review for cases with low R-scores.

\section{Data and Methods}
\label{sec:data-methods}

We evaluate our framework on a text annotation task of classifying scientific papers by their primary contribution type: 
\emph{Interesting Hypothesis}, 
\emph{Technical Advance}, 
or \emph{New Finding}. 
Below we summarize the data, models, and how we implement both the survey-inspired interventions and independent probability assessment. We also describe how we test the downstream effects of unreliable annotations in a citation-impact regression.

\subsection{Data}
\paragraph{F1000 Dataset.}
The study utilized the F1000 dataset (also known as Faculty Opinions) from previous studies in science of science for demonstration. It comes from a post-publication peer-review platform in which invited scholars — practicing scientists and clinicians — select and evaluate biomedical papers they deem significant. The experts are asked to label papers with predefined tags in about five categories. For demonstration purposes, the labels used here contain three primary contribution types that takes the majority of cases: (A). Interesting Hypothesis (7.5\%)
(B). Technical Advance (13.3\%) and (C). New Finding (79\%). These expert annotations has been shown to align with different types of novelty \citep{shi2023surprising}, thus are important in studying science and innovation. 

This dataset is suitable here as it provides expert-validated classifications that requires nuanced understanding. The categories are distinct yet related, making it a moderately challenging test for LLM reliability. Notably, the class imbalance mirrors real-world scientific output - most papers make empirical findings over theoretical or methodological contributions. This distribution creates natural test conditions for evaluating LLM reliability across frequent and rare categories. After preprocessing and cleaning, a total of 816 biomedical papers with expert annotations are included. 

\paragraph{Microsoft Academic Graph (MAG).}
To examine downstream effects, this study also uses a simple example of predicting a paper's citation impact (with in 3 years of publication) based on the paper's contribution types using linear regression models. To do so, the F1000 data is merged with Microsoft Academic Graph Dataset, which has publicly available Microsoft Academic Graph, to get the citation count of 816 papers. The merge was done by matching their MAG paper ids with their PMID in the PubMed dataset. Citation counts follow a heavy-tailed distribution (mean=142, SD=213, max=2,184), typical of scientific impact patterns. We log-transform citations after adding 1 to handle zeros. We also added year and team size as control variables.

\subsection{LLM Models}
To facilitate analysis, open-sourced LLMs are favored for analysis. The study employs three variants of the leading models -- LLaMA-3.1 Instruct series (8B, 70B, and 405B parameters) to systematically examine how model annotation reliability changes under survey-inspired interventions. The LLaMa series capture the spectrum from lightweight to state-of-the-art LLMs (open sourced), allowing us to test whether larger models exhibit greater robustness to survey-inspired interventions. All models use the standard dense decoder-only autoregressive transformer architecture along with supervised fine-tuning and direct preference optimization after pretraining \citep{meta2024llama3}.
The tripartite model selection directly informs key findings in later analyses: 1) The 8B model serves as a baseline for "commodity" LLMs accessible on consumer hardware, showing high intervention sensitivity (Figure \ref{fig:fig3} - \ref{fig:fig5}) due to limited contextual reasoning capacity. 2) The 70B variant represents current practical limits of dense models, demonstrating partial robustness to position randomization but remaining vulnerable to reverse validation (Figure \ref{fig:fig3} - \ref{fig:fig5}). 3) The 405B model tests whether large models with extreme scale can overcome satisficing tendencies - the results suggest even this frontier model retains non-trivial sensitivity to option ordering (Figure \ref{fig:fig3} - \ref{fig:fig5}), indicating the importance of case-level reliability assessment for current LLM paradigms.

To ensure comparability across model sizes, the analysis maintains identical generation parameters: temperature=0 for controlled randomness, top-p=0.7 sampling, and maximum output length=1 token). Change of parameters don't affect the main outcomes. For probability distribution analysis, the study extracts logits directly from the final unembedding layer by using TogetherAI api with the parameter "logprobs" equals True . This setup allows us to precisely track how intervention-induced perturbations affect the models' internal confidence metrics at the precise token of interest.

\subsection{Survey-Inspired Interventions}
\label{ssec:interventions}
The study apply s\textbf{option randomization}, \textbf{position randomization}, and \textbf{reverse validation} to each of the 816 papers. Each intervention has multiple variants. In particular, the paper measures the \emph{flip rate}, i.e., the proportion of cases where the model’s top label changes relative to the original prompt. This yields a detailed diagnostic of how each model size responds to small prompt changes.

\subsection{Independent Probability Assessment and R-score}
\label{ssec:indep-assessment}
The study then implements the \emph{independent binary query} strategy for the same papers. For example, it asks: 
- ``Is the main contribution of this paper an \emph{Interesting Hypothesis}?'' 
- ``Is the main contribution of this paper a \emph{Technical Advance}?'' 
- ``Is the main contribution of this paper a \emph{New Finding}?''

From these three queries, the study extracts $p(\text{Yes}|c_i)$ for $i \in \{\text{A, B, C}\}$, normalize them to $p(c_i)$, and compute the Kullback-Leibler divergence from uniform random guessing, $R = D_{KL}(P||U)$. Using thresholds, the study categorize each annotation's reliability from ``very low'' to ``high.'' 

\subsection{Downstream Regression Analysis}
Finally, the study demonstrates how these reliability issues can alter analytic conclusions, by regressing $\log(\text{citations}+1)$ on an indicator variable for (B) Technical Advance (versus C) and observe how coefficient estimates shift when using labels produced under various interventions. Significant changes in sign or significance underscore the risk of ignoring order sensitivity and reliability in real-world social science studies.

\section{Results}
\label{sec:results}

\subsection{Diagnosing Order Sensitivity via Flip Rates}
\begin{figure}[H]
\centering
\includegraphics[width=0.65\textwidth]{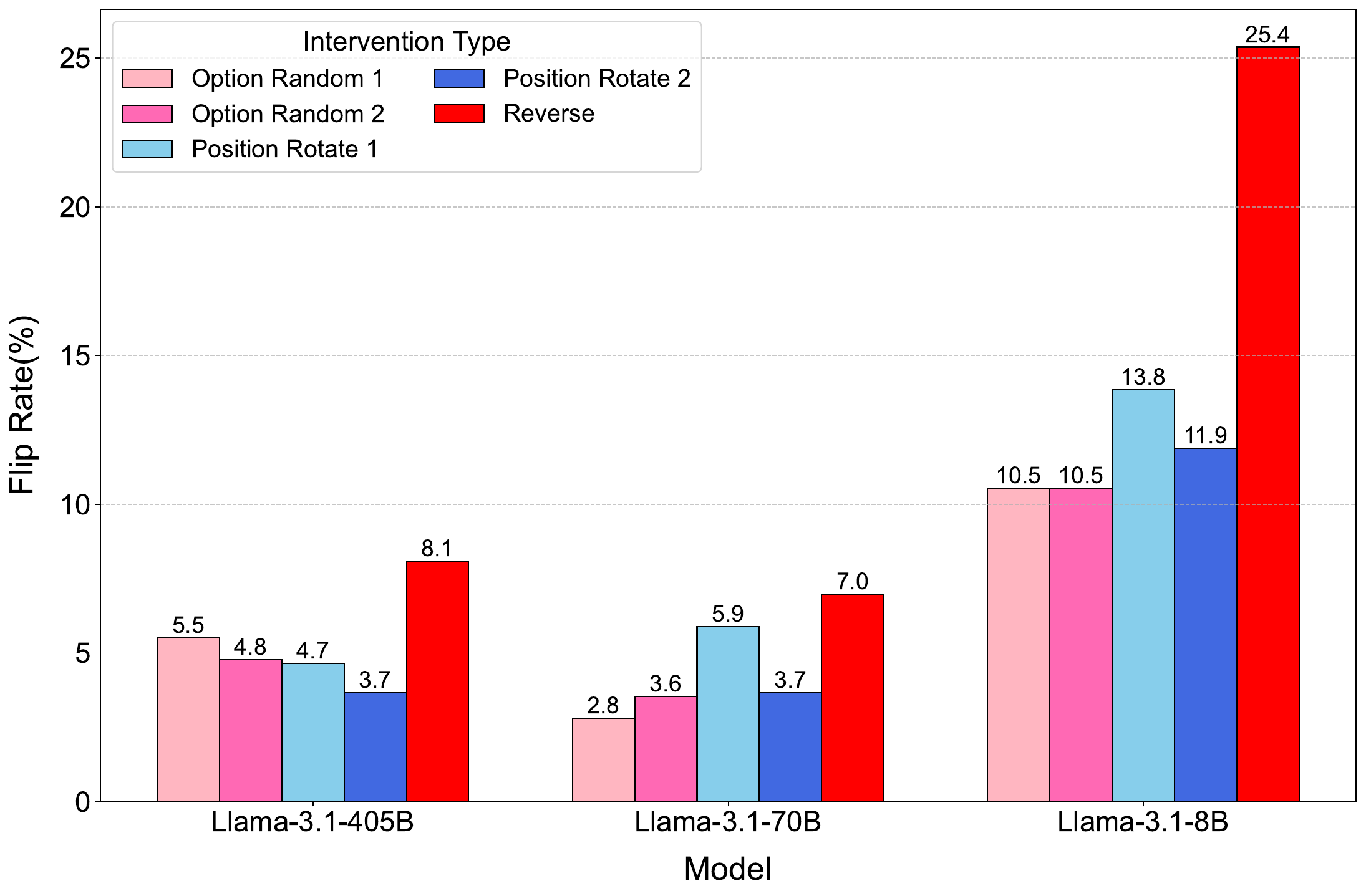}
\caption{\textbf{Flip Rates under Survey-Inspired Interventions.} Percentage of instances where the model’s top label changes when we randomize label options (pink), randomize prompt position (blue), or apply reverse-coded logic (red). Data shown for LLaMA-3.1 with 8B, 70B, and 405B parameters. Smaller models exhibit higher flip rates, indicating strong order sensitivity, but even the largest model is not fully immune.}
\label{fig:fig3}
\end{figure}

Figure \ref{fig:fig3} reports the overall \emph{flip rates} for the three interventions across our three LLaMA-3.1 models. Even the largest (405B) model exhibits non-negligible flips (5--10\%) across different intervention types, while the 8B model can flip in over 25\% of cases under some variants. These results demonstrate that the fundamental problem of \emph{order sensitivity} persists for large models, though it is more pronounced for smaller ones. 

\begin{figure}[H]
\centering
\includegraphics[width=1\textwidth]{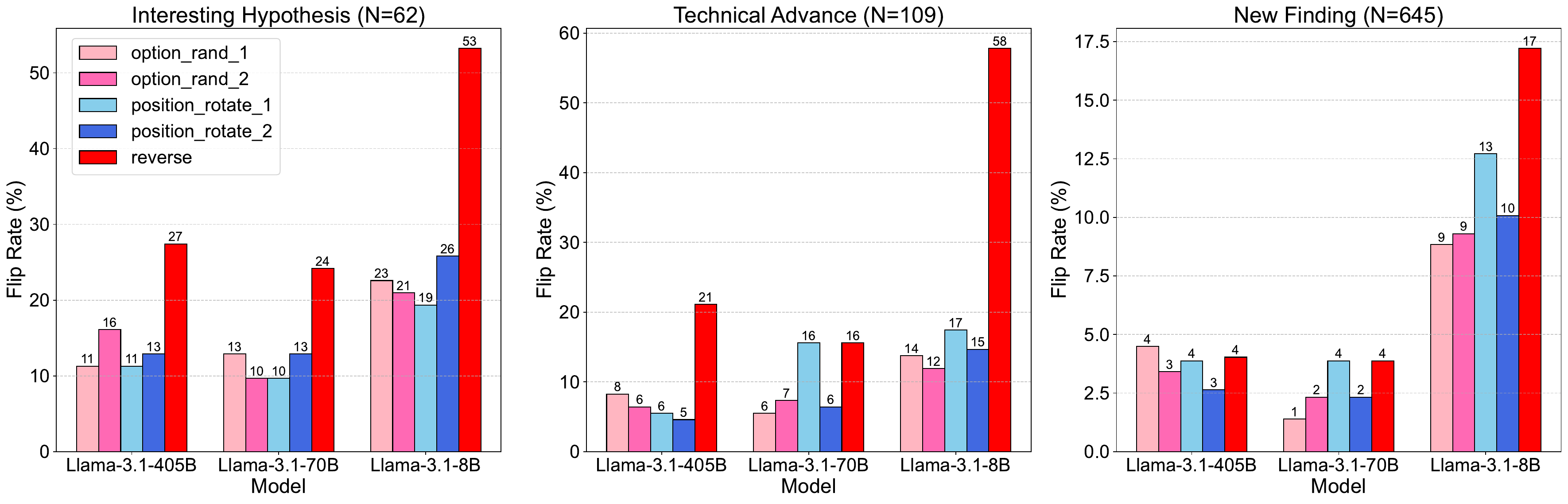}
\caption{\textbf{Flip Rates by Category.} Depending on whether the paper is labeled by experts as (A) Interesting Hypothesis (N=62), (B) Technical Advance (N=109), or (C) New Finding (N=645), flip rates vary under option randomization, position randomization, and reverse validation. Rare categories (like Interesting Hypothesis) are disproportionately affected.}
\label{fig:fig4}
\end{figure}

Figure \ref{fig:fig4} zooms in on how flip rates vary by paper category -- Interesting Hypothesis (N=62), Technical Advance (N=109), and New Finding (N=645) -- across all three LLMs. A key takeaway is that the “Interesting Hypothesis” category shows the highest flip rates in most interventions for all model sizes. This suggests that rarer or more conceptually demanding classes may induce greater model uncertainty, making them more susceptible to small perturbations in option ordering, prompt position, or question framing. In other words, the models struggle disproportionately with classifying the less common or more abstract paper types, indicating a particular instability that could significantly affect downstream analyses focused on such minority categories. This instability in underrepresented cases will not be identified if researchers soling relying on the external metrics (such as accuracy) for model evaluation.

For “Technical Advance,” the flip rates also remain elevated, especially for the 8B model, but somewhat moderate for the 70B and 405B models. “New Finding,” by contrast, shows lower but still non-zero flip rates; this is likely due to its dominance in the training distribution and, hence, the model’s learned preference for that label. Nonetheless, the models still register flips in the “New Finding” category under certain interventions (especially reverse validation). 

\subsection{Consistency vs. Accuracy}

\begin{figure}[H]
\centering
\includegraphics[width=1\textwidth]{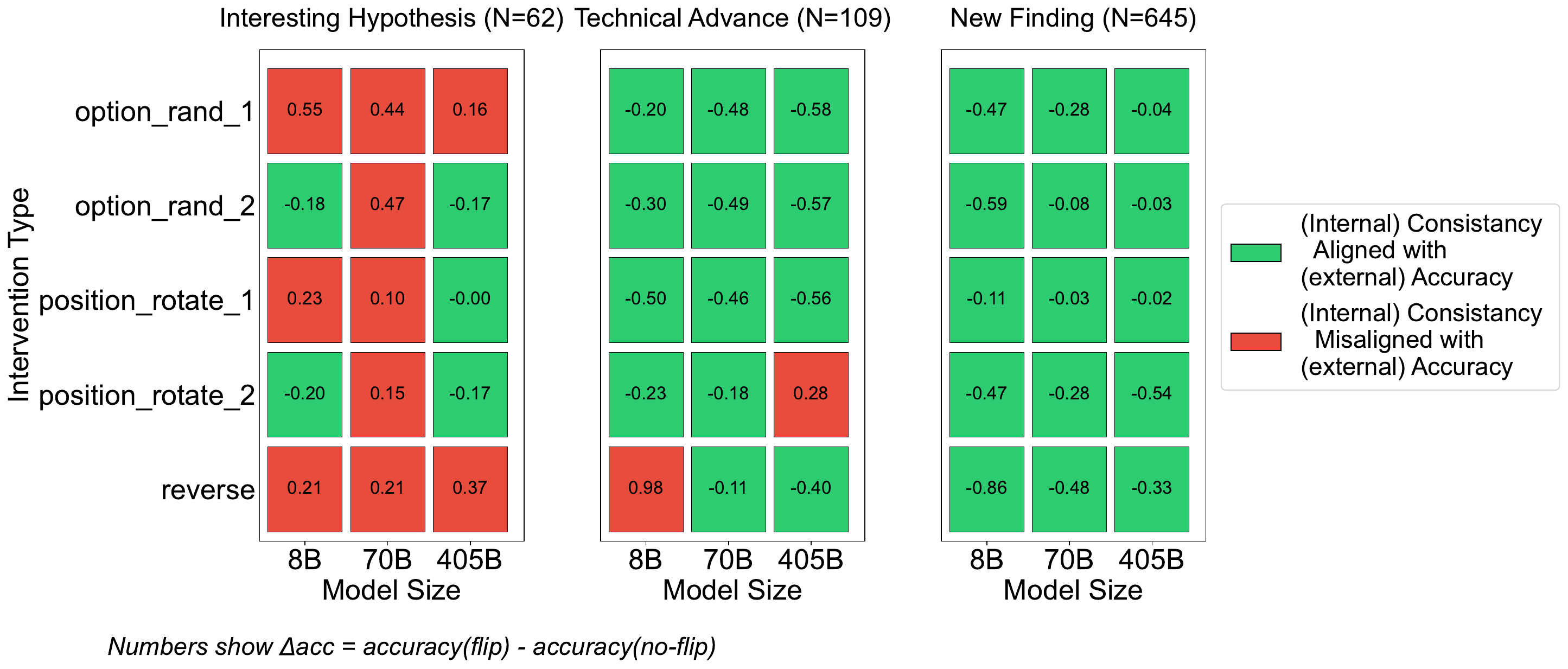}
\caption{\textbf{Consistency vs. Accuracy ($\Delta acc$).} Each matrix cell captures $\Delta acc$: the difference in accuracy between flipped vs. non-flipped cases. Red indicates higher accuracy for flipped subsets; green indicates higher accuracy for stable subsets. Patterns reveal that flipping is not always negatively correlated with accuracy, complicating naive assumptions that consistent answers are always better.}
\label{fig:fig5}
\end{figure}

Figure \ref{fig:fig5} explores how ``internal consistency” (whether an annotation flips under an intervention) aligns—or fails to align—with “external accuracy” (agreement with expert ground truth). Each cell reports $\Delta acc$, defined as accuracy(flip) minus accuracy(no-flip). By design, a positive $\Delta acc$ (shown in red) means that those annotations which flipped under an intervention ironically ended up being more accurate than those that stayed the same. Conversely, negative $\Delta acc$ (green) indicates that flips tend to be bad indicators for correctness in those cases -- i.e., changing an answer correlates with lower accuracy relative to not flipping. 
The result shows that only the ``New Finding" category shows alignment between external accuracy and internal consistancy for all interventions. In the other categories, all three models show misalignment in certain situations. This finding carries significant implications: although conventional wisdom suggests that higher internal consistency should correspond to higher accuracy and thus better for use, these positive $\Delta acc$ values in minor categories (especially in ``Interesting Hypothesis") shows that this assumption does not always hold. The areas of discrepancy consistute a ``non-sense region" in which the model is unreliable, and in which simple reliance on external validation alone can be misleading.

\clearpage
\subsection{Independent Probability Assessment: R-score Distributions}

\subsubsection{Examples of Unreliable Annotations}

\begin{table}[htbp]
    \centering
    \caption{Example of Unreliable LLM Annotation with Very Low Reliability Score}
    \begin{tabular}{|p{3cm}|p{9cm}|}
        \hline
        \multicolumn{2}{|c|}{\textbf{Paper Information}} \\
        \hline
        \textbf{Title} & Setting health care priorities in Oregon. Cost-effectiveness meets the rule of rescue \\
        \hline
        \textbf{Abstract} & The Oregon Health Services Commission recently completed work on its principal charge: creation of a prioritized list of health care services, ranging from the most important to the least important. Oregon's draft priority list was criticized because it seemed to favor minor treatments over lifesaving ones. This reaction reflects a fundamental and irreconcilable conflict between cost-effectiveness analysis and the powerful human proclivity to rescue endangered life: the "Rule of Rescue." Oregon's final priority list was generated without reference to costs and is, therefore, more intuitively sensible than the initial list. However, the utility of the final list is limited by its lack of specificity with regard to conditions and treatments. An alternative approach for setting health care priorities would circumvent the Rule of Rescue by carefully defining necessary indications for treatment. Such an approach might be applied to Oregon's final list in order to achieve better specificity. \\
        \hline
        \multicolumn{2}{|c|}{\textbf{Annotation Comparison}} \\
        \hline
        \textbf{Expert Annotation} & Interesting Hypothesis \\
        \hline
        \textbf{LLM Annotation} & \textit{Llama-3.1-8B probability distribution (assessed independently and then normalized):} \\
        & Interesting Hypothesis: 0.326 \\
        & Technical Advance: 0.337 \\
        & New Finding: 0.337 \\
        \hline
        \textbf{Reliability Score} & 0.00 (Very Low Reliability) \\
        \hline
        
        \hline
    \end{tabular}
    \label{tab:unreliable_example}
\end{table}

Using the Reliability-score, the study was able to identify at the case level which annotations are unreliable. Below I present two of the unreliable cases from the F1000 dataset. One annotation of the two cases is inaccurate compared to expert annotation and have low reliablility score, and another one is accuracy but still have low reliability. 

The first example shown in Table \ref{tab:unreliable_example} demonstrates a case of low reliability in LLM annotation (Inaccurate and Unreliable). The LLaMa-3.1 8B model assigns nearly identical probabilities to all three categories (approximately 1/3 each), indicating it cannot meaningfully distinguish between them for this paper. The resulting Reliability Score of 0 signifies that the probability distribution is effectively uniform, equivalent to random guessing. Despite this uncertainty, if forced to choose, the model would incorrectly label this as either "Technical Advance" or "New Finding" rather than the expert-assigned "Interesting Hypothesis" category. This case illustrates how accuracy metrics alone would flag this as an incorrect annotation, but the proposed reliability score additionally reveals that the model has no meaningful confidence in its answer.

\clearpage
\begin{table}[htbp]
    \centering
    \caption{Example of Low Reliablilty LLM Annotation}
    \begin{tabular}{|p{3cm}|p{9cm}|}
        \hline
        \multicolumn{2}{|c|}{\textbf{Paper Information}} \\
        \hline
        \textbf{Title} & Dynamics of age-structured and spatially structured predator-prey interactions: individual-based models and population-level formulations. \\
        \hline
        \textbf{Abstract} & In this article, we investigate the spatial and temporal dynamics of predator and prey populations using an individual-based modeling approach. In our models, the individual is the fundamental unit, and the dynamics are governed by individual rules for growth, movement, reproduction, feeding, and mortality. We first establish the congruence between age-structured predator-prey population models and the corresponding individual-based population model under homogeneous spatial conditions. Given the agreement between the formalisms, we then use the individual-based model to investigate the dynamics of spatially structured predator-prey systems. In particular, we contrast the dynamics of predator-prey systems in which predators adopt either an ``ambush'' or a ``cruising'' strategy. We show that the stability of the spatially structured predator-prey system depends on the relative mobility of prey and predators and that prey mobility, in particular, has a strong effect on stability. Local density dependence in prey reproduction can quantitatively alter the asymmetrical influence of prey mobility on stability, but we show that the asymmetry exists when local density dependence is removed. We hypothesize that this asymmetrical response is due to prey ``escape'' in space caused by differences in rates of spread of prey and predator populations that arise because of fundamental differences between prey and predator reproduction. \\
        \hline
        \multicolumn{2}{|c|}{\textbf{Annotation Comparison}} \\
        \hline
        \textbf{Expert Annotation} & New Finding \\
        \hline
        \textbf{LLM Annotation} & \textit{Llama-3.1-405B probability distribution(assessed independently and then normalized):} \\
        & Interesting Hypothesis: 0.117 \\
        & Technical Advance: 0.384 \\
        & New Finding: 0.498 \\
        \hline
        \textbf{Reliability Score} & 0.13 (Low Reliability) \\
        \hline
        
    \end{tabular}
    \label{tab:moderate_reliability_example}
\end{table}

The second example in Table \ref{tab:moderate_reliability_example} demonstrates a different case of low reliability in LLM annotation (Correct but Still Unreliable). While the model correctly identifies ``New Finding'' as the most probable category (0.498), there remains substantial uncertainty. The model assigns a significant probability (0.384) to ``Technical Advance,'' indicating it struggles to fully differentiate between these categories for this paper. The reliability score of 0.13 suggests the model is slightly confident in its classification but still exhibits notable uncertainty. This case illustrates a common challenge in scientific paper classification: the boundary between reporting new empirical findings and introducing novel technical approaches can be subtle, particularly in computational modeling papers like this one. Traditional accuracy metrics would simply mark this as correct, missing the important nuance that the model's confidence is relatively low.

\begin{figure}[H]
\centering
\includegraphics[width=1\textwidth]{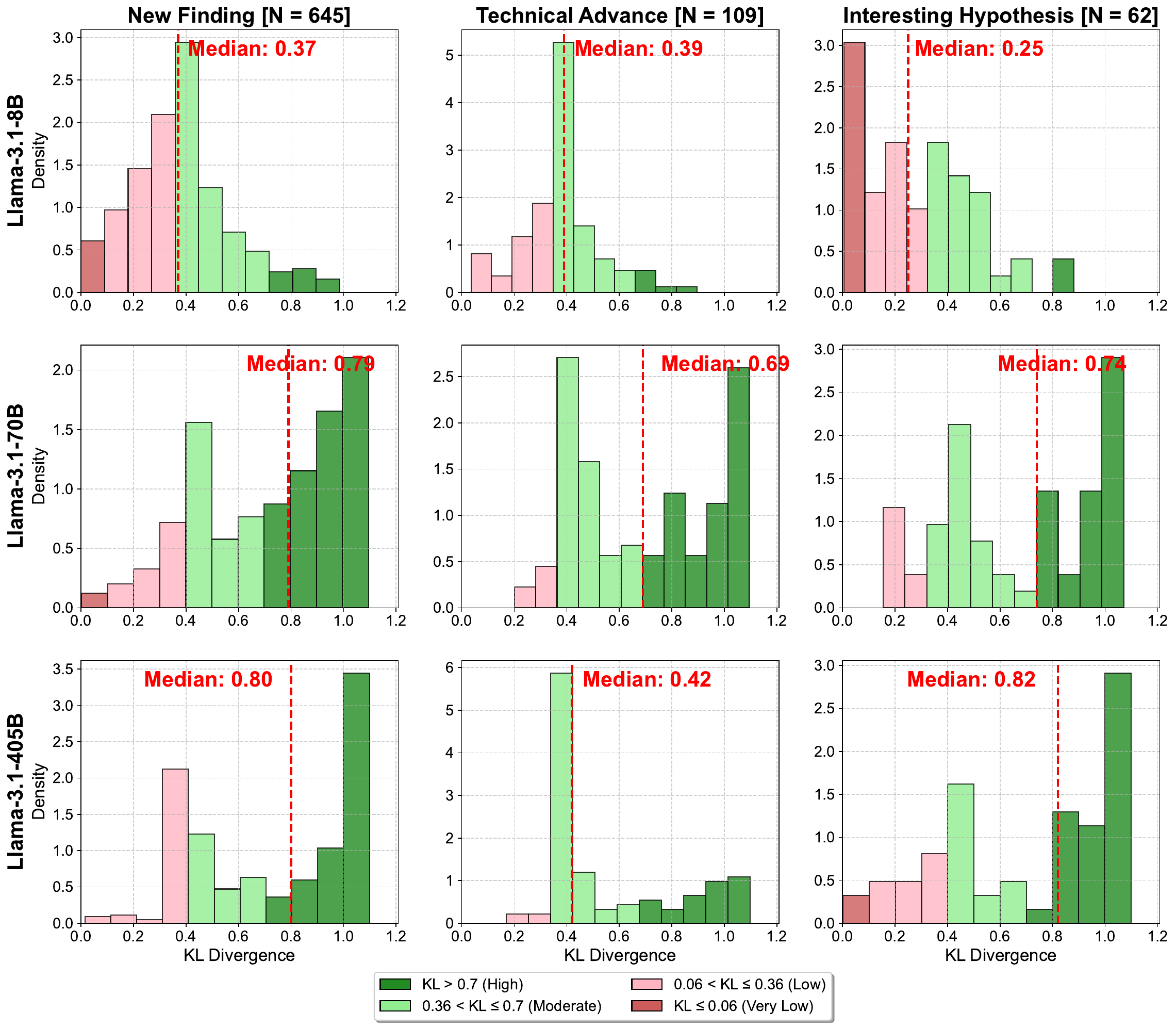}
\caption{\textbf{R-score Distribution by Model and Category.} For each paper, the study applies the independent binary query method to obtain a probability distribution and compute KL divergence from uniform. Higher values (green) indicate high reliability (strong preference for a single label), while lower values (red) suggest the model is essentially guessing. Although larger models exhibit generally higher reliability, some categories (especially rarer ones) remain problematic.}
\label{fig:fig6}
\end{figure}

The next part shows how the \emph{independent probability assessment} addresses the causal ordering bias. Figure \ref{fig:fig6} presents the resulting \emph{R-score} distributions across the three category labels and three model sizes. Higher R-scores indicate the model confidently distinguishes among categories, while lower scores suggest a near-uniform (random) distribution.

Smaller (8B) models yield a considerable fraction of near-uniform distributions, reflecting fundamental uncertainty. Larger models show improved reliability overall, but still produce many low or moderate R-scores for rare categories like \emph{Interesting Hypothesis}. While larger models generally demonstrate higher overall reliability (median KL of 0.37, 0.79, and 0.80 for 8B, 70B, and 405B respectively in the New Finding category), there is notable variation in reliability across different annotation categories within the same model. For instance, the 405B model shows high reliability for New Finding (median KL = 0.80) and Interesting Hypothesis (median KL = 0.82) but only moderate reliability for Technical Advance (median KL = 0.42). This variability shows that even when a model demonstrates high overall reliability, its performance may still be notably unreliable for specific categories or individual cases, as identified by R-score.

\subsection{Relation Between Flip Rates and R-score}
Although omitted the figure in this paper for space, empirical results also confirm that cases with low R-scores (below 0.06, for example) are the same ones most likely to flip under label reordering or reverse-coded prompts. This aligns with the interpretation that if the model’s \emph{internal distribution} is nearly uniform, small changes to the prompt context can easily shift the top label. In contrast, a high R-score indicates robust preference that is resilient to ordering artifacts.

\subsection{Downstream Impact on Regression Findings}
The paper next illustrates the practical impact of ignoring order sensitivity in research by regressing (log) citation count on the LLM-predicted label of whether a paper is a \emph{Technical Advance} (vs.\ the baseline \emph{New Finding}). Table~\ref{tab:regression_results} portrays how sensitive the coefficient estimates can be when the classification depends on prompts with slightly different label orders or reversed logic. In some conditions, the same LLM that produced a non-significant coefficient in the original prompt yields a highly significant and negative coefficient after a prompt change. Such discrepancies are especially large for smaller models. These findings reveal how ignoring order sensitivity can jeopardize the robustness of conclusions about scientific impact or other social science outcomes.

\begin{table}[!htbp]
\centering
\caption{Impact of Different Interventions on Downstream Regression Tasks}
\label{tab:regression_results}
\small
\begin{tabular}{p{3.2cm}ccp{1.2cm}cp{1.2cm}c}
\toprule
& & & \multicolumn{2}{c}{Original Classification} & \multicolumn{2}{c}{Intervention Classification} \\
\cmidrule(lr){4-5} \cmidrule(lr){6-7}
Intervention Type & N & $R^2$ & Coef. & p-value & Coef. & p-value \\
\midrule
\multicolumn{7}{l}{\textbf{Panel A: 8B Model}} \\
\midrule
\rowcolor{red!20} Option Random 1 & 615 & 0.171 & -0.060 & 0.615 & -0.322 & 0.010\sym{**} \\
Option Random 2 & 615 & 0.171 & -0.060 & 0.615 & -0.071 & 0.594 \\
\rowcolor{red!20} Position Rotate 1 & 615 & 0.171 & -0.060 & 0.615 & -0.255 & 0.023\sym{**} \\
Position Rotate 2 & 615 & 0.171 & -0.060 & 0.615 & 0.045 & 0.739 \\
\rowcolor{red!20} Reverse & 615 & 0.171 & -0.060 & 0.615 & -1.206 & 0.005\sym{***} \\
\midrule
\multicolumn{7}{l}{\textbf{Panel B: 70B Model}} \\
\midrule
\rowcolor{red!20} Option Random 1 & 612 & 0.178 & -0.446 & 0.009\sym{***} & -0.268 & 0.076\sym{*} \\
Option Random 2 & 612 & 0.178 & -0.446 & 0.009\sym{***} & -0.396 & 0.010\sym{***} \\
\rowcolor{red!20} Position Rotate 1 & 612 & 0.178 & -0.446 & 0.009\sym{***} & -0.341 & 0.026\sym{**} \\
\rowcolor{red!20} Position Rotate 2 & 612 & 0.178 & -0.446 & 0.009\sym{***} & -0.322 & 0.041\sym{**} \\
\rowcolor{red!20} Reverse & 612 & 0.178 & -0.446 & 0.009\sym{***} & -0.071 & 0.690 \\
\midrule
\multicolumn{7}{l}{\textbf{Panel C: 405B Model}} \\
\midrule
Option Random 1 & 619 & 0.169 & -0.189 & 0.196 & -0.127 & 0.346 \\
\rowcolor{red!20} Option Random 2 & 619 & 0.169 & -0.189 & 0.196 & -0.268 & 0.060\sym{*} \\
Option Random 1 & 619 & 0.169 & -0.189 & 0.196 & -0.185 & 0.202 \\
Option Random 2 & 619 & 0.169 & -0.189 & 0.196 & -0.100 & 0.520 \\
Reverse & 619 & 0.169 & -0.189 & 0.196 & 0.038 & 0.793 \\
\bottomrule
\multicolumn{7}{p{13cm}}{\small \textit{Notes:} Dependent variable is the log of journal citations within 3 years of publication. All regressions include year and team size fixed effects.} \\[0.2em]
\multicolumn{7}{p{13cm}}{\small \textit{Coefficients} show the effect of Technical Advance (as compared to New Finding) on citation counts. Highlighted rows indicate significant changes in statistical significance level of coefficients.} \\[0.2em]
\multicolumn{7}{p{13cm}}{\small \sym{*}$p<0.1$, \sym{**}$p<0.05$, \sym{***}$p<0.01$. Heteroskedasticity-robust standard errors (HC1).}
\end{tabular}
\end{table}

\section{Discussion}
\label{sec:discussion}
The results underscore that causal transformers’ sequential processing is a fundamental source of order sensitivity in text classification prompts. The survey-inspired interventions show that reordering labels, altering the position of the question, or posing reverse-coded queries can induce substantial volatility in the predicted label. Critically, this volatility remains hidden if researchers rely solely on a single prompt design and measure accuracy against that one scenario.

Beyond diagnosing the problem, the \textbf{independent probability assessment} offers a technical remedy for multi-choice tasks: it elicits category probabilities one at a time, thereby circumventing the causal architecture’s positional bias. By looking at the final distribution of those probabilities via our \textbf{R-score}, we can further identify cases that are near-random guesses versus cases where the model expresses strong confidence. In practice, social scientists can use this approach to:

\begin{enumerate}[leftmargin=*]
    \item \textbf{Filter or weight LLM annotations by R-score.} Exclude low-R-score instances or treat them as uncertain to avoid spurious signals.
    \item \textbf{Target expert validation.} Reserve time-consuming human reviews for the suspicious or borderline subset of cases, potentially focusing on rare categories where the model is known to be least reliable.
    \item \textbf{Pretest prompts with survey-like interventions.} Identify if order sensitivity is acute for a given classification task before deploying on large datasets, mitigating unpredictable label noise.
\end{enumerate}

\subsection{Relation to Survey Satisficing and Annotation Tasks Broaderly}
The findings parallel issues in survey methodology, where inattentive or ``satisficing'' respondents are detected by randomizing question orders or including reverse-coded items \citep{krosnick1991response, barge2012using}. Similarly, LLMs sometimes appear to select shortcuts based on positional or lexical cues instead of deeply engaging the content. While the root cause in LLMs is grounded in the mechanics of causal attention rather than human cognitive effort, the outcomes---unstable or low-fidelity responses---are reminiscent of survey data quality concerns.

These results are helpful for social science research that seeks to harness LLMs at scale. Rare categories, such as novel theoretical constructs, are often the focus of interest in specialized domains, yet they are precisely where these reliability issues emerge most strongly. Researchers adopting LLM-based annotation pipelines should thus be especially cautious with imbalanced or conceptually nuanced labels.

\subsection{Limitations and Future Directions}
One limitation of the approach is the added computational overhead of running multiple independent queries (one per category). Another is that while the framework illustrate the power of order-sensitive interventions and the R-score for classification tasks with a moderate number of categories, extensions to tasks with dozens or hundreds of labels would require further efficiency considerations. Moreover, other aspects of LLM outputs (e.g., chain-of-thought reasoning, open-ended text generation) may exhibit forms of sensitivity not addressed here. Future work could explore how to adapt R-score-like measures to open-ended tasks or how to integrate the independent assessment framework deeper into model pre-training.

\section{Conclusion}
\label{sec:conclusion}
Different technologies follow their own logics \citep{mackenzie1999social}.

LLM-based annotation, much like the financial modelling practices \citep{mackenzie2014formula}, operates not in a vacuum of pure computation but within a specific technical and methodological context -- what might be termed an emerging "computational annotation culture" within social science. This culture leverages powerful tools, predominantly causal transformer models, initially developed for tasks like text generation where their core architectural properties -- sequential processing and causal attention -- are precisely what enable their remarkable fluency and coherence.

However, as this study has demonstrated, this very architecture imposes a fundamental structural constraint when these tools are repurposed for tasks demanding position-invariant evaluation, such as multi-choice classification. The sequential processing of labels and prompt elements introduces a systemic order sensitivity: the model's internal evaluation and probability assignment for a given category can become unnervingly dependent on its arbitrary position within the input sequence, rather than solely on the substantive content relative to the category definition. This isn't a mere bug; it's a feature of the causal design manifesting as a significant vulnerability when deployed in a context requiring robust, context-free (relative to prompt structure) judgment.

This research underscores that relying solely on traditional accuracy metrics, which measure agreement against a single ground truth under one specific prompt design, provides an incomplete and potentially misleading picture of an annotation pipeline's true performance. Accuracy in this context often captures a form of "local correctness" relative to a single setup, but it fails to reveal the underlying brittleness and order-dependence that can undermine the "global reliability" needed for robust social science inference.

To navigate this challenge, the paper proposes a methodological "bricolage," adapting the practices to the specific constraints of the tool. Old experience in survey research helps. The survey-inspired interventions -- option randomization, position randomization, and reverse validation -- serve as crucial diagnostic probes. Akin to the stress tests developed in other domains, they reveal the extent of this hidden order sensitivity through empirical "flip rates." The findings show that this fragility is not uniform; it is often acutely pronounced for rarer or conceptually more nuanced categories, precisely where social scientists often focus their analytical attention. This highlights a significant gap between the perceived sophistication of LLM outputs and their actual reliability for specific, demanding annotation tasks. 

Furthermore, the proposed independent probability assessment offers a technical adaptation specifically designed to mitigate the causal ordering bias in commonly used multi-choice settings. By disentangling the evaluation of each category into separate queries, it effectively bypasses the causal attention mechanism's tendency to compare options based on their sequence. This process allows for eliciting a more faithful representation of the model's internal confidence distribution, without the burden of heavy internal analysis of model parameters and activations \citep{li2023inference, kim2025linear, zou2023representation}. The derived R-score then translates this distribution into a case-level measure of reliability, allowing researchers to identify and manage the underlying uncertainty at the granularity required for robust social scientific analysis. A low R-score indicates the model is near-random guessing, a critical signal missed by a simple top-label prediction.

Implementing such methods is not without its trade-offs, echoing the interplay between goals, resources, and constraints observed in other technical fields. Running multiple independent queries inherently increases computational demands compared to a single multi-choice prompt. This mirrors the material constraints (hardware capacity, processing time, energy budgets) that shaped the adoption and adaptation of models like the Gaussian copula in financial institutions. What makes an LLM annotation model truly "work" for social science is not simply conceptual elegance or the highest score on a decontextualized benchmark, but its ability to produce stable, interpretable, and reliably uncertain signals within the practical constraints and analytical objectives of the research. The framework provides tools to make these trade-offs and the resulting reliability explicit.

Finally, the demonstration of the downstream impact on regression analyses serves as a stark warning. Building analytical frameworks and drawing substantive conclusions based on annotations susceptible to these subtle, architecturally-induced shifts can lead to findings that are themselves unstable, inconsistent, or even contradictory. Just as financial models enabling "local stability" in specific trading desks could mask systemic risks contributing to broader market instability, seemingly "accurate" LLM annotations (when evaluated naively) can mask an underlying fragility that undermines the validity of broader scientific claims derived from them. Relying on these powerful computational tools without a rigorous methodology that acknowledges their specific architectural properties, diagnoses their vulnerabilities, and quantifies their uncertainty is akin to building a theoretical structure on an unstable foundation. It is crucial for social scientists to move beyond the seductive fluency and apparent ease of LLM-based annotation and embrace methodological approaches that explicitly account for the ways in which these tools interact with the data and the research task, ensuring that digital annotations provide a robust and reliable basis for understanding the social world.


\end{document}